\documentclass[pss,fleqn]{w-art}
\usepackage{times}
\usepackage{amssymb}
\usepackage{w-thm}
\usepackage[]{graphicx}
\begin{document}
\DOIsuffix{theDOIsuffix}
\Volume{XX}
\Issue{X}
\Month{XX}
\Year{2004}
\pagespan{3}{}
\Receiveddate{May 2004}
\Reviseddate{XXX}
\Accepteddate{XXX}
\Dateposted{XXX}



\title[]{Influence of surface anisotropy on the magnetization reversal of nanoparticles}


\author[\`O. Iglesias]{\`Oscar Iglesias \footnote{Corresponding
     author: e-mail: {\sf oscar@ffn.ub.es}, Phone: +34\,93\,4021155,
     Fax: +34\,93\,4021149}\inst{1}} \address[\inst{1}]{Departament de \ F\'{\i}sica Fonamental, Universitat de Barcelona, Diagonal 647, 08028 Barcelona, Spain}
\author[A. Labarta]{Am\'{\i}lcar Labarta\inst{1}}
\begin{abstract}
The influence of surface anisotropy on the magnetization processes of maghemite nanoparticles with ellipsoidal shape is studied by means of Monte Carlo simulations. Radial surface anisotropy is found to favor the formation of hedgehog-like spin structures that become more stable as the surface anisotropy constant at the surface $k_S$ is increased form the value at the core. We have studied the change in the low temperature hysteresis loops with the particle aspect ratio and with $k_S$, finding a change in the magnetization reversal mode as $k_S$ or the particle elongation is increased.
\end{abstract}
\maketitle                   





\section{Introduction}
Most spins in a magnetic nanoparticle are at the particle surface where the reduced coordination and the broken translational symmetry of the lattice are responsible for changes in some local physical properties of the material. Among them, there is the reduction in exchange interaction constants and the increased local anisotropy at the particle surface with respect to bulk values \cite{Tartajprb04} phenomenological model that described this last phenomenon was first introduced by Ne\'el in \cite{Neeljpr51}, but experimental determination of the microscopic structure and properties of the surface of nanoparticles is still a challenge to be solved. For example, an estimate of the surface anisotropy constant can only extracted indirectly from magnetic measurements by fitting to an effective anisotropy constant \cite{Bodkerprl94}. However, there are increasing evidences from experiments that surface effects play a dominant role in the anomalous low temperature magnetic properties of nanoparticle systems \cite{Batllejpd02}. Therefore, simulations of magnetic properties in which microscopic parameters extracted form experiments are used \cite{Trohidoujap98,Kachkachiprb02,Labayjap02} may serve as a benchmark to test the phenomenological models at hand. In this article, we use the Monte Carlo method to simulate the magnetization processes of individual nanoparticles modeled by classical Heisenberg spins on the maghemite lattice with radial surface anisotropy $k_S$ different from that on the core. 

\section{Model and simulation method}
The model considered is an extension to Heisenberg spins of our previous MC simulation for an Ising spin model of a $\gamma$-Fe$_2$O$_3$ (maghemite) particle \cite{Iglesiasprb01}. The particles considered have spherical or ellipsoidal shape with diameter $D= 3$ and elongations $L= 3-8$ (in units of the cell size $a$), containing from $N= 347$ to $883$ spins. The surface spins of the particle are considered to be those on the outer unit cell and have anisotropy constant $k_S$ and radial local anisotropy axes along the directions $\bf{n}_i$, whereas core spins have uniaxial anisotropy $k_C$ pointing along the z-axis. Indirect estimations of the anisotropy constants yield values in the range $k_C= 0.01-1$ K, $k_S= 1- 100$ K, so we will fix $k_C= 1$ K and vary only $k_S$. The interaction Hamiltonian in temperature units can be written as 
\begin{eqnarray}
\label{Hamiltonian}
{ H}/k_{B}= 
-\sum_{\langle  i,j\rangle}J_{ij} {\vec S}_i \cdot {\vec S}_j   
-\sum_{i= 1}^{N} \vec h\cdot{\vec S_i}
-\sum_{i= 1}^{N_{Core}}\left[k_C(S_i^z)^2\right]+\sum_{i= 1}^{N_{Surf}}\left[k_S(\vec{S}_i \cdot \hat n_i)^2 \right]\ . 	
\end{eqnarray}
In the first term, $J_{ij}$ are the nearest-neighbours exchange constants among the spins in the two sublattices with different coordination of maghemite, all of them are antiferromagnetic, being $J_{TO}$ the strongest in magnitude and making maghemite a ferrimagnetic compound \cite{Kodamaprb99}. We have computed equilibrium configurations by starting from a disordered state at a high temperature and decreasing the temperature in steps $dT= 1$. Hysteresis loops are started from a saturating field along the core easy-axis that is decreased in steps  $dh= -1$, during which the magnetization is averaged over $1000$ MC steps. In both cases, the standard MC method for continuous spins has been used with a trial jump consisting in a combination of uniform and small angle around the current spin direction that allows to obtain reliable simulations for treat the weak and strong anisotropy cases with the same algorithm, see \cite{IglesiasphaB04} for the details. 
\begin{figure}[h]
\centering
\caption{Equilibrium spin configurations of an ellipsoidal maghemite particle with radial surface anisotropy, diameter $D= 3a$ and elongation $L= 8a$, achieved after cooling form high temperature to $T=0$. Upper (lower) panels are cross-sections along the equatorial plane (long-axis) showing only the spins on the central unit cell. 
} 
\label{Fig_1}
\end{figure}

\section{Results}
In Fig. \ref{Fig_1}, we show snapshots of the spin configurations attained after the cooling process previously described, for several values of the surface anisotropy constant $k_S$. The pictures present a cut of one unit cell width along the equatorial plane and along the major axis of the particle. For values of $k_S$ smaller than a certain critical value $k_S^{\star}$ (which depends on the particle size and on the shape), the radial anisotropy at the surface is not strong enough to cant the core spins towards the radial direction. Therefore, the core spins are antiferromagnetically aligned with a small net magnetization component along the $z$ axis due a small number of non-compensated spins (see Fig. \ref{Fig_1} for $k_S= 10$). However, at the particle surface, the spins present already a high degree of disorder due to the competition between the reduced local exchange field (locally reduced with respect to that on the core spins) and the anisotropy, inducing some degree of frustration at the surface. As $k_S$ is further increased (see Fig. \ref{Fig_1} for $k_S =20$), the disorder of the surface spins is transmitted to the particle core and results in the departure of core spins from the local anisotropy axis towards the radial direction. Finally, when anisotropy dominates over exchange energy ($k_S > k_S^{\star}$, see Fig. \ref{Fig_1} for $k_S= 100$), the hedgehog-like structures are formed, with the spins pointing along the local radial direction and having antiparallel orientations with the nearest neighbors on the other sublattice. 

\begin{figure}[h]
\centering
\includegraphics[width= 1.0\columnwidth]{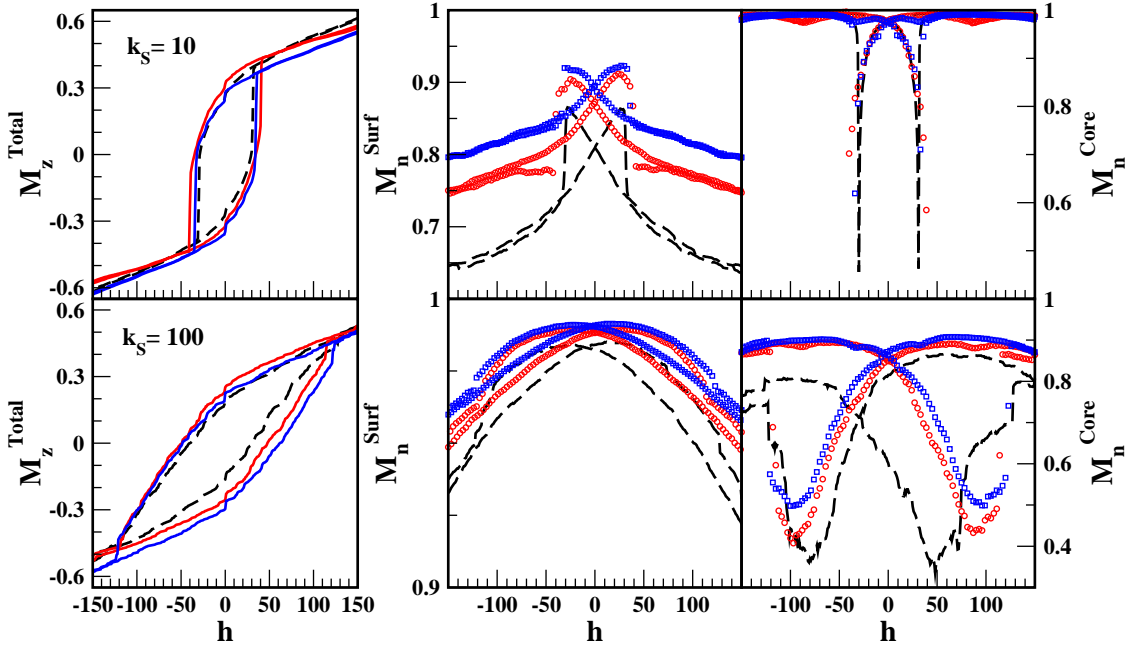}
\caption{Hysteresis loops for ellipsoidal particles with two values of the radial surface anisotropy constant $k_S= 10, 100$. Dashed lines are for a spherical particle with diameter $D= 3$, red circles and blue squares for an ellipsoidal particle with $D= 3$ and $L= 6,8$ respectively. Left panels display the total magnetization while central and right panels show the surface and core contributions to $M_n$, the sum of projections of the magnetization in to the local easy-axis directions.
} 
\label{Fig_2}
\end{figure}

We have also studied the influence of surface anisotropy and particle shape on the magnetization reversal processes. For this purpose, zero temperature hysteresis loops have been simulated using the procedure described previously for ellipsoidal particles with elongations $L= 3-8$ and diameter $D= 3$. Some selected results are presented in Fig. \ref{Fig_2}, where the total magnetization component along the field direction (left panels) is shown together with the contribution of the surface and core spins to $M_{n}=\sum_{i=1}^N \left| \vec S_i\cdot \hat n_i \right|$ , the sum 
of the projections of spins onto the local anisotropy axis (central and right panels). By comparing the loops for $k_S$= 10 to those for $k_S$= 100, we see that the coercive field increases and the remanence decreases with increasing surface anisotropy, independently of the elongation L of the particle. Moreover, the presence of disordered groups of spins at the surface induced by surface anisotropy makes the loops more elongated and increases the closure fields of the loops as found also in experiments on ferrimagnetic oxides \cite{Martinezprl98}. The rounding of the loops near the coercive field clearly indicates a progressive departure from a uniform reversal mechanism with increasing $k_S$. When looking only at the $M_z$ component, not much difference is appreciated between the loops for particles with different L because of the compensation of the spin components transverse to the field direction due to the symmetry of revolution of the particles around the $z$ axis. However, upon further inspection of the $M_n^{Surf}$ and $M_n^{Core}$ contributions and animated snapshots taken along the loops \cite{WWW_SCM2004}, the details of the reversal process can be better under-stood. Let us notice first that, when $k_S$ is increased from $k_S^{\star}$, the reversal mechanism changes from quasi-uniform (induced by the core) rotation at low $k_S$ values, to a process in which the formation of surface hedgehog-like structures induce the non-uniform switching of the whole particle. In the first regime ($k_S= 10$ case in Fig. \ref{Fig_2}), the core and surface spins point mostly along the z axis ($M_n^{Core} \sim 1$, $M_n^{Surf} \ll 1$) except near the coercive field, where they make short excursions towards the radial direction driven by the surface anisotropy (see the dips in $M_n^{Core}$ and the cusps in $M_n^{Surf}$). However, for $k_S$> $k_S^{\star}$ ($k_S= 100$ case in Fig. \ref{Fig_2}), surface spins remain near the local radial easy-directions ($M_n^{Surf} \sim 1$) during the reversal, while the core spins are dragged out the z local easy-axis by the surface spins during the reversal, except for values of h near the closure field. This is indicated by the widening of the dips in $M_n^{Core}$ and the global decrease of $M_n^{Core}$ values as $k_S$ increases. Finally, let us remark also that, for all the $k_S$ considered, the $M_n^{Surf}$ values along the whole hysteresis loops increase with increasing $L$, which indicates that the surface spins remain closer to the local radial direction during the reversal as particles become more elongated. When increasing $L$, the dips in $M_n^{Core}$ become less profound for $k_S > k_S^{\star}$, and indication that reversal of core spins along the radial direction is suppressed by the elongation. However, for weak anisotropy ($k_S < k_S^{\star}$), the more elongated the particles are, the greater the deviation of surface spins towards the radial direction during the reversal.

\begin{acknowledgement}
  We acknowledge CESCA and CEPBA under coordination of C$^4$ for computer facilities.
  This work has been supported by the spanish SEEUID through the MAT2003-01124 project
  and the Generalitat de Catalunya through the 2001SGR00066 CIRIT project.
\end{acknowledgement}

\end{document}